\documentclass{mem}
\usepackage{natbib}\usepackage{txfonts}\usepackage{balance}
\usepackage{graphicx}
\usepackage[a4paper,breaklinks,dvipdfm]{hyperref}
\idline{0}{1}
\begin{document}

\title{
An intermediate-band photometric study of the ``Globular Cluster'' NGC~2419
}

\author{
Andreas \, Koch\inst{1}
\and
Matthias \, Frank\inst{1} 
\and
Sofia \, Feltzing\inst{2} 
\and
Daniel \, Ad\'en\inst{2}
\and\\
Nikolay \, Kacharov\inst{1} 
\and 
Mark I. \, Wilkinson\inst{3}
}

  \offprints{A. Koch}

\institute{
Zentrum f\"ur Astronomie der Universit\"at Heidelberg,  Landessternwarte, K\"onigstuhl 12, 69117 Heidelberg, Germany. 
\email{akoch@lsw.uni-heidelberg.de}
\and
Lund Observatory, Box 43, SE-22100 Lund, Sweden
\and
Department of Physics and Astronomy, University of Leicester, University Road, Leicester LE1 7RH
}

\authorrunning{A. Koch et al.}

\titlerunning{Str\"omgren photometry of NGC~2419}

\abstract{
NGC 2419 is  one of the remotest star clusters in the Milky Way halo 
and its exact nature is yet unclear: 
While it has traits reminiscent of a globular cluster, its large radius and 
suggestions of an abundance spread have fueled the discussion about its origin 
in an extragalactic environment, possibly the remnant of 
the accretion of a dwarf galaxy.  
Here, we present first results from deep intermediate-band photometry of NGC 2419, which enables us to  
search for chemical (light element) abundance variations, metallicity spreads, and thus multiple stellar populations through well calibrated Str\"omgren 
indices.  
}
\maketitle{}
\section{Introduction}
\subsection{NGC 2419}
NGC 2419 is a stellar aggregate with a number of puzzling characteristics and its nature and origin are yet unclear. 
With a half-light radius $r_h$ of 21.4 pc it is the fifth-most extended object listed in the 2010-version of the Harris (1996) catalogue, while 
it is also one of the most luminous Globular Clusters (GCs) in the Milky Way (MW; Fig.~1). At a Galactocentric distance of 90 kpc it 
resides in the outermost halo. All these traits have fueled discussions of whether it contains any dark matter or could be affected by non-Newtonian
dynamics (Baumgardt et al. 2005; Conroy et al. 2011; Ibata et al. 2012).  For instance,  Ibata et al. (2012) argue that its kinematics is incompatible with a dark matter content in excess of some 6\% of its total mass. 
Overall, these morphological and dynamical considerations beg the question to what extent NGC~2419 has evolved in isolation and 
whether it could be associated with a once-accreted, larger system like a dwarf (spheroidal) galaxy.
\begin{figure}[t!]
\resizebox{\hsize}{!}{
\includegraphics[clip=true]{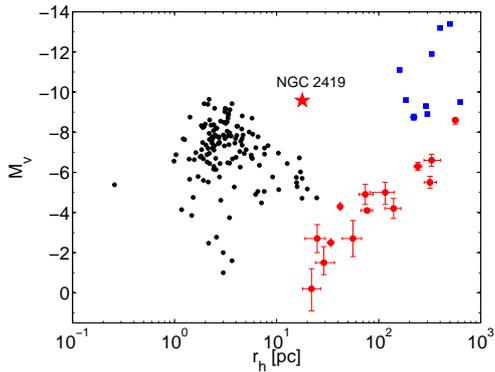}
}
\caption{\footnotesize
Magnitude-half light radius plot for GCs (black dots), luminous dSphs (blue squares) and ultrafaint MW satellites (red circles). NGC~2419 is labeled -- what is this object?}
\vspace{-0.5cm}
\end{figure}

Also chemically, NGC 2419 has much to offer: Cohen \& Kirby (2012) and Mucciarelli et al. (2012) identified a population of stars (ca. 30\% by number) with remarkably low Mg- and high K-abundances, 
which could be the result of ``extreme nucleosynthesis'' (Ventura et al. 2012). The question of an abundance-spread has been addressed by several authors using high-resolution spectroscopy and 
low-resolution measurements of the calcium triplet. However,   the large abundance variation of 
the electron-donor Mg will upset the commonly used stellar model atmospheres so that any claimed spread in iron-,  Ca-, and thus overall  metallicity needs to be considered with caution. 
However, settling exactly this aspect is of prime importance, since any significant spread in heavy elements is a trademark signature of an object with a likely extragalactic origin 
(e.g., Fig.~1 in Koch et al. 2012). 

The color-magnitude diagrams (CMDs) of  Di Criscienzo et al. (2011) show a hint of a color-spread towards the subgiant branch and the presence of a 
hot, faint Horizontal Branch (HB), 
consistent with a second generation of stars with a strongly increased He-content. Thus, also NGC~2419 does appear to show signs of multiple stellar populations,
in line with the majority of the MW GC system. 
\subsection{Str\"omgren photometry}
While broad-band filter combinations have succeeded in unveiling multiple stellar populations in sufficiently deep data sets and more massive systems (e.g., Piotto et al. 2007), additional 
observations in intermediate-band 
Str\"omgren filters
are desirable for a number of reasons:
\begin{enumerate}
\item[\em i)] The $c_1 = (u-v)- (v-b)$ index in combination with a color such as $v-y$ is a powerful {\em dwarf/giant separator} and can efficiently remove any foreground contamination (e.g., Faria et al. 2007). 
At $b=25\degr$ this can be expected to be  less of a problem in NGC~2419, but see, e.g., Ad\'en et al. (2009) for an impressive demonstration of such a CMD cleaning. 
Our first assessment of the $c_1$-$(b-y)$ plane indicates that the foreground contamination is indeed minimal on the upper RGB (see also Fig.~2). 
\item[\em ii)] The index $m_1 = (v-b)- (b-y)$ is a good proxy for stellar {\em metallicity} and calibrations have been devised by several authors (e.g., Hilker 2000; Calamida et al. 2007; 
Ad\'en et al. 2009). 
\item[\em iii)] {\em Multiple populations} in terms of split red giant branches (RGBs), multiple subgiant branches, and main sequence turnoffs are well separated in CMDs that use combinations 
of Str\"omgren filters, e.g., $\delta_4 = c_1 + m_1$ (Carretta et al. 2011), where optical CMDs based on broad-band filters still show unimodal, ``simple stellar populations''. 
\item[\em iv)] This is immediately interlinked with the {\em chemical abundance variations} in the light chemical elements (e.g., Anthony-Twarog et al. 1995) that accompany the multiple populations, 
most prominently driven by N-variations. Accordingly, Yong et al. (2008) confirmed linear correlations of  $c_y = c_1 - (b-y)$ with the [N/Fe] ratio.
\end{enumerate}
\section{Data and analysis}
We obtained imaging in all relevant Str\"omgren filters ($u$,$b$,$v$,$y$) using the Wide Field Camera (WFC) at the 2.5-m Isaac Newton Telescope (INT) at La Palma, Spain. 
Its large field of view ($33\arcmin\times33\arcmin$) allows us to trace the large extent of NGC~2419 out to several times its tidal radius ($r_t\sim7.5\arcmin$). 

Instrumental magnitudes were obtained via PSF-fitting using the \textsc{Daophot/Allframe} software packages 
(Stetson 1987).
The instrumental magnitudes were transformed to the standard Str\"omgren system using ample observations of standard stars 
(Schuster \& Nissen 1988).
We set up transformation equations similar to those given by Grundahl, Stetson \& Andersen (2002).
\section{Preliminary results: CMDs and [M/H]}
Fig.~2 shows two CMDs of NGC~2419, where we restrict our analysis to the bona-fide region between 1 and 3 half-light radii to avoid potentially crowded, inner regions, yet 
minimizing the field star contamination of the outer parts. 
For the present analysis, we adopted a constant reddening of E($B-V$)$= 0.061$\,mag\footnote{Obtained from \url{http://irsa.ipac.caltech.edu/applications/DUST}}, 
and its respective transformations to the Str\"omgren system (Calamida et al. 2009).
\begin{figure*}[t!]
\resizebox{\hsize}{!}{
\includegraphics[width=0.02\hsize]{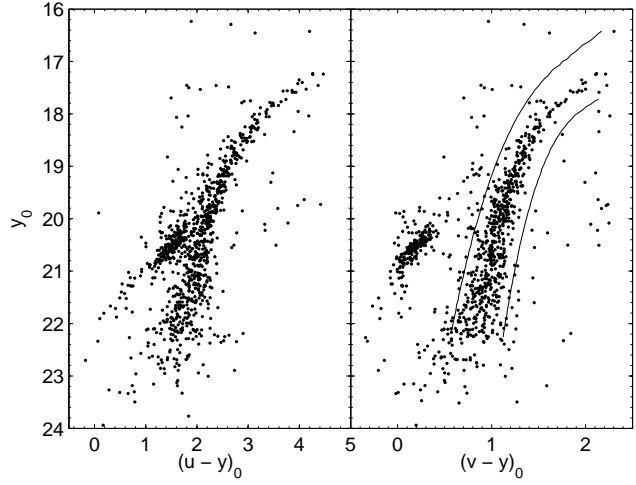}
}
\caption{\footnotesize
CMDs of NGC 2419 for two possible Str\"omgren-filter combinations. Shown are stars between $r_h < r <  3\,r_h$; no other selection criteria have been applied. 
Stars falling within the lines shown right  were used to construct the metallicity distribution in Fig.~3.}
\end{figure*}

While we do not resolve the main sequence turnoff, our photometry reaches about 1 mag below the HB at $y_{\rm HB}$$\equiv$$V_{\rm HB}$$\sim$$20.5$ mag. 
All regions of the CMD are well reproduced, showing 
a clear RGB, hints of an AGB and bright AGB (which stand out more clearly in other color indices; Frank et al. in prep.), and a prominent HB. Hess diagrams also highlight the presence of a clear 
RGB bump at $y_0\sim20.3$ mag. 
Moreover, the extreme, hotter HB stands out in the bluer u-band (left panel), confirming the presence of this  He-rich, secondary population (di Crisicienzo et al. 2011). 

To obtain a first impression of the metallicity distribution function (MDF)  of NGC~2419 (Fig.~3, right panel), we convert our Str\"omgren photometry to metallicities, [M/H], through the 
calibration by Ad\'en et al. (2009). 
This was carried out for stars on the RGB (see ridge lines in Fig.~2, right panel, and Fig.~3, left). As a result, we find 
a mean [M/H] of $-2$ dex. This is  in good agreement with the values listed in the Harris catalogue and the high-resolution data of Mucciarelli et al. (2012) and Cohen \& Kirby (2012) of [Fe/H] = $-2.15$ dex. 
\begin{figure*}[t!]
\resizebox{\hsize}{!}{
\includegraphics[clip=true,width=0.55\hsize]{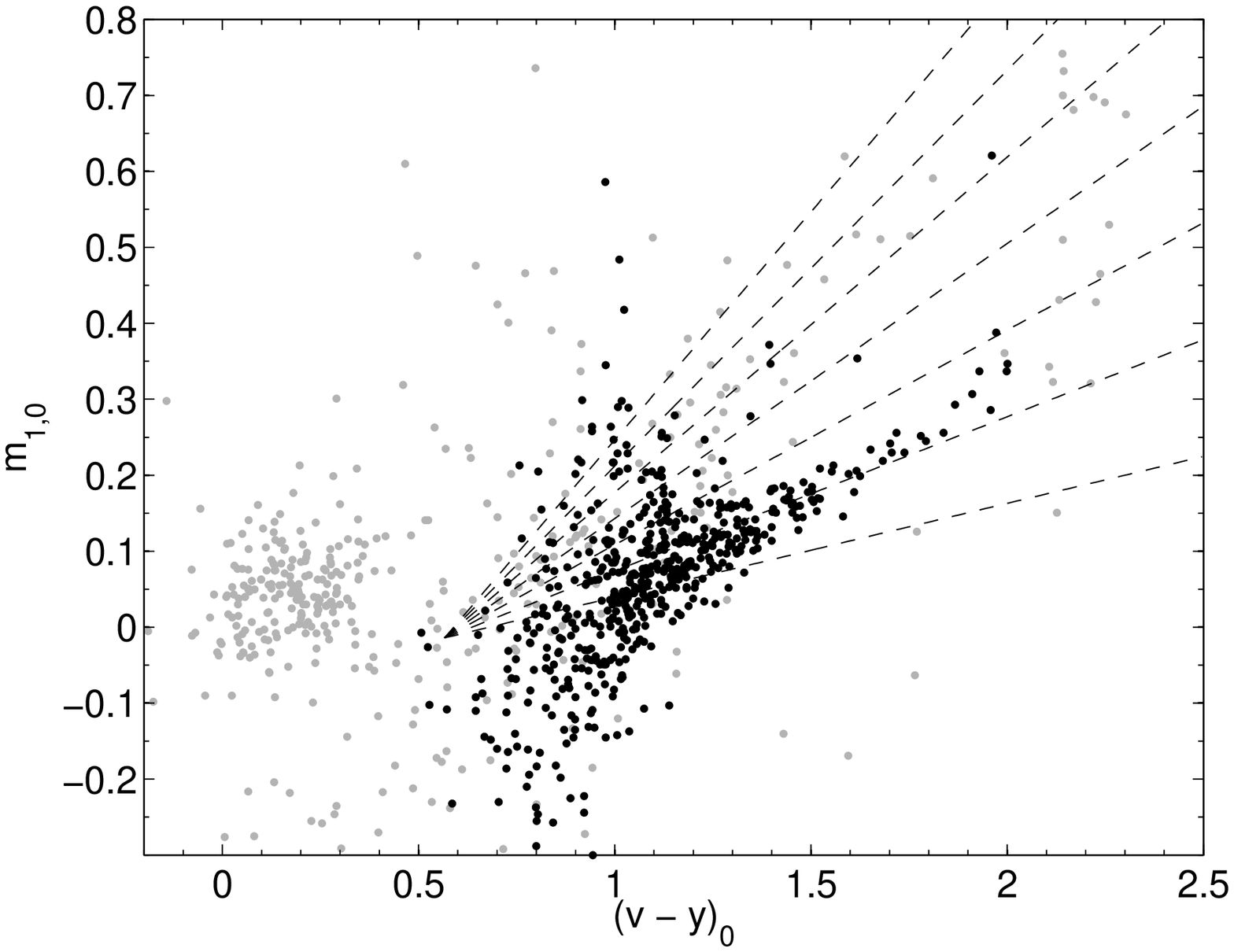}
\includegraphics[clip=true,width=0.53\hsize]{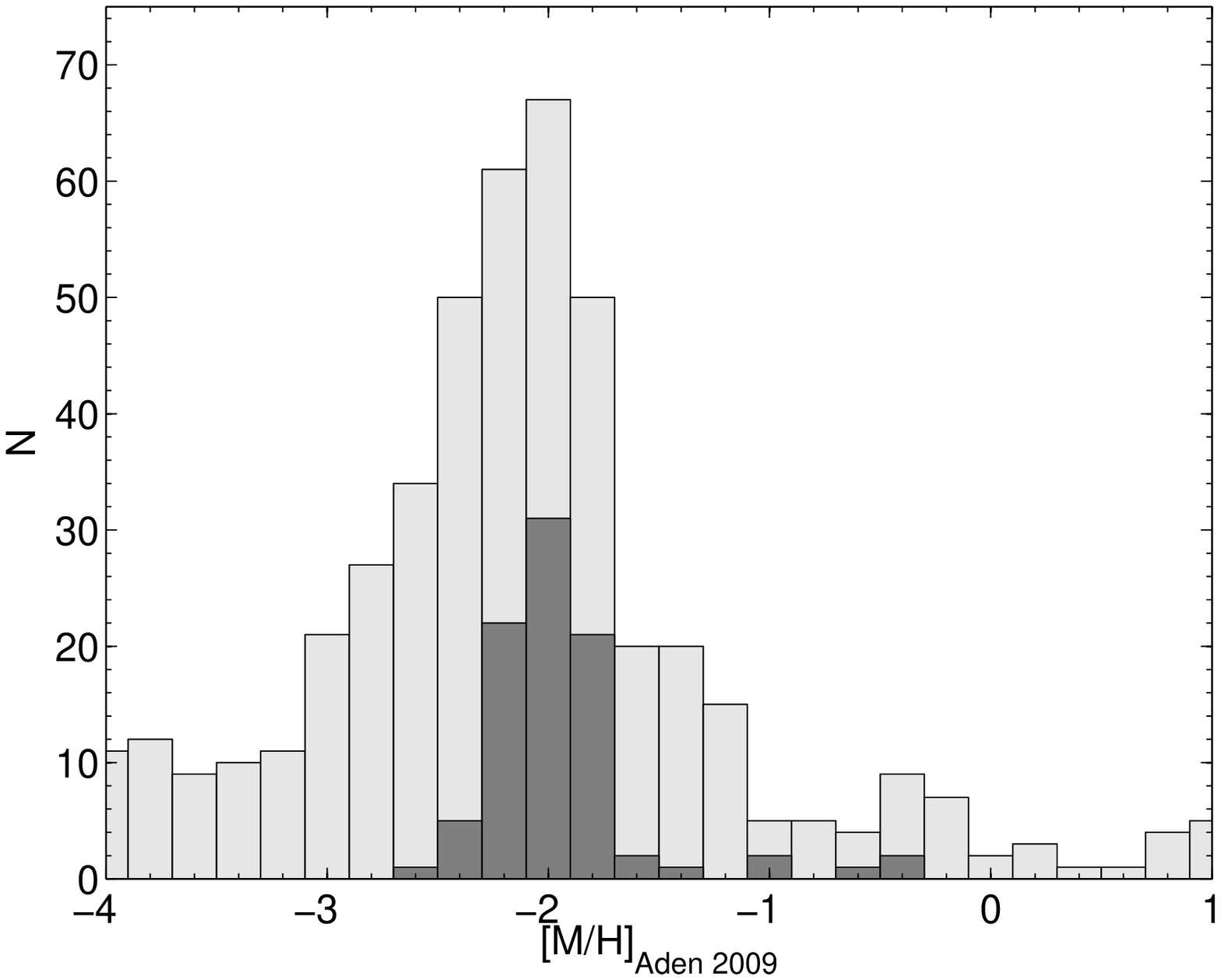}
}
\caption{\footnotesize
Preliminary metallicity calibration in the $m_1$-vs-$(b-y)$ plane (left panel). Dashed lines indicate iso-metallicity curves based on the calibration of Ad\'en et al.  (2009) for [M/H] = $-2.5$ up to +0.5 dex in steps of 0.5 (bottom to top). 
Black dots are those within the RGB-ridge lines of Fig.~2, used to infer the MDF (light gray) in the right panel. The dark gray MDF in this plot uses a stricter RGB criterion of $(v-y)_0 > 1.4$.}
\label{eta}
\end{figure*}

The MDF also indicates the presence of a broad metallicity spread, where we find a nominal 1$\sigma$-spread of 0.5 dex, but this is probably still dominated by 
remaining foreground contaminants and photometric errors. While we cannot exclude the presence of an abundance spread in NGC 2419 from the 
present data, it is very likely much smaller than the one suggested by Fig.~3.
\section{Discussion} 
Although our CMD does not allow us to clearly isolate any multiple stellar populations at this stage of our analysis,  we nevertheless   
find strong reason to believe in their presence in NGC~2419, bolstered by recent optical images (di Criscienzo et al. 2011). 
These authors detected a color spread at the base of the RGB and an extreme, hot HB, indicative of an increased He-abundance of a populous second stellar generation.  
This HB population is also visible in our intermediate-band CMDs.  

Although our first analysis suggests a broad metallicity spread in NGC~2419, this is probably not significant and further CMD filtering is necessary. 
However, our derived mean metallicity is in line with the results from high-resolution spectroscopy, which indicates that  our Str\"omgren photometry 
is well calibrated. 
\begin{acknowledgements}
AK, MF, and NK gratefully acknowledge the Deutsche Forschungsgemeinschaft for funding from  Emmy-Noether grant  Ko 4161/1. 
This research has made use of the NASA/ IPAC Infrared Science Archive, which is operated by the Jet Propulsion Laboratory, California Institute of 
Technology, under contract with the National Aeronautics and Space Administration.
\end{acknowledgements}
\bibliographystyle{aa}

\begin{thebibliography}{}
%
\bibitem[Ad{\'e}n et al.(2009)]{2009A&A...506.1147A} Ad{\'e}n, D., Feltzing, S., Koch, A., et al.\ 2009, \aap, 506, 1147 
%
\bibitem[Anthony-Twarog et al.(1995)]{1995PASP..107...32A} Anthony-Twarog, B.~J., Twarog, B.~A., \& Craig, J.\ 1995, \pasp, 107, 32 
%
\bibitem[Baumgardt et al.(2005)]{2005MNRAS.359L...1B} Baumgardt, H., et al.  \ 2005, \mnras, 359, L1 
%
\bibitem[Calamida et al.(2007)]{2007ApJ...670..400C} Calamida, A.,  et al.\ 2007, \apj, 670, 400 
%
\bibitem[Calamida et al.(2009)]{2009ApJ...706.1277C} Calamida, A.,  et al.\ 2009, \apj, 706, 1277
%
\bibitem[Carretta et al.(2011)]{2011A&A...535A.121C} Carretta, E., et al.\ 2011, \aap, 535, A121 
%
\bibitem[Cohen \& Kirby(2012)]{2012ApJ...760...86C} Cohen, J.~G., \& Kirby, E.~N.\ 2012, \apj, 760, 86 
%
\bibitem[Conroy et al.(2011)]{2011ApJ...741...72C} Conroy, C., et al.\ 2011, \apj, 741, 72 
%
\bibitem[di Criscienzo et al.(2011)]{2011MNRAS.414.3381D} di Criscienzo, M., D'Antona, F., Milone, A.~P., et al.\ 2011, \mnras, 414, 3381 
%
\bibitem[Faria et  al.(2007)]{2007A&A...465..357F} Faria, D., Feltzing, S., Lundstr{\"o}m, I., et al.\ 2007, \aap, 465, 357 
%
\bibitem[Grundahl et al.(2002)]{2002A&A...395..481G} Grundahl, F., et al. \ 2002, \aap, 395, 481
%
\bibitem[Harris(1996)]{1996AJ....112.1487H} Harris, W.~E.\ 1996, \aj, 112, 1487 
%
\bibitem[Hilker(2000)]{2000A&A...355..994H} Hilker, M.\ 2000, \aap, 355, 994 
%
\bibitem[Ibata et al.(2012)]{2012MNRAS.tmp..225I} Ibata, R.,  et al.\ 2012, \mnras, 225 
%
\bibitem[Koch et al.(2012)]{2012EPJWC..1903002K} Koch, A., et al. \ 2012, European Physical Journal Web of Conferences, 19, 3002 
%
\bibitem[Mucciarelli et al.(2012)]{2012MNRAS.426.2889M} Mucciarelli, A., Bellazzini, M., Ibata, R., et al.\ 2012, \mnras, 426, 2889 
%
\bibitem[Piotto et al.(2007)]{2007ApJ...661L..53P} Piotto, G.,  et al.\ 2007, \apjl, 661, L53 
%
\bibitem[Schuster \& Nissen(1988)]{1988A&AS...73..225S} Schuster, W.~J., \& Nissen, P.~E.\ 1988, \aaps, 73, 225 
%
\bibitem[Stetson(1987)]{1987PASP...99..191S} Stetson, P.~B.\ 1987, \pasp, 99, 191
%
\bibitem[Ventura et al.(2012)]{2012ApJ...761L..30V} Ventura, P., D'Antona, F., Di Criscienzo, M., et al.\ 2012, \apjl, 761, L30 
%
\bibitem[Yong et al.(2008)]{2008ApJ...684.1159Y} Yong, D., Grundahl, F., Johnson, J.~A., \& Asplund, M.\ 2008, \apj, 684, 1159 
%
%
\end{thebibliography}
\end{document}